# Fast and Modular Analog Rank Order Filter Using CMOS Technology

R. Mukund and Amit Kumar Mishra[1]

Department of Computer Science and Engineering,
[1]Electronics and Communication Engineering, Indian Institute of Technology, Guwahati - 781 039, Assam, India

**ABSTRACT**

In this paper, we apply the sorting network theory to construct an analog rank order filter. We present a voltage mode CMOS two-input sorting element and arrange these to form a rank order filter. The resulting circuit is simple and is a high-speed, high-precision design. Although the transistor count is moderately higher than other designs, the proposed circuit simultaneously outputs signals of all ranks rather than of just one specified rank. We also develop a slightly modified design which calculates the rank of a given signal. We present reports of simulations to verify the performance of the device.

## 1. INTRODUCTION

Extracting the minimum, maximum and median of a set of signals is an important activity in signal processing. More generally, the $k^{th}$ order value, when a set of signals is arranged in ascending order, is desired [1,2]. A device calculating this value is termed a rank order filter (ROF). ROFs are an important class of nonlinear filters often implemented in software. This necessitates the use of analog to digital converters (ADC) and hence slows down signal processing. The use of analog ROFs enables us to shift this part of the signal processing operation to hardware and increases throughput. Hardware complexity is also reduced since we can select the data to pass to the later processing stages. Several authors have developed analog rank order filters [3-7]. Previous approaches include using transconductance amplifiers and artificial neural networks.

As signal processing circuits grow larger, it is desirable to reduce the transistor count in the modules. A high-speed ROF with $O(n)$ transistor count has been reported [4]. However, if we wished to simultaneously determine signals of multiple ranks for a given set of inputs, we would have to replicate the whole circuit. In the worst case, where we want to sort the signals in order, $O(n^2)$ transistors would be required. A good solution to this problem will be useful when we report maximum, minimum and median of a given set of signals. The design presented in this paper would reduce this to $O(n \log n)$ transistors.

It is often desirable to perform the inverse of the ROF operation-answering the question, "How many signals in the input have higher magnitude than this one?" This involves finding the rank of a given signal. Existing methods do not lend them to extension in this manner.

An analog ROF may be implemented as either a voltage or current mode design. We present a voltage mode design. Another characteristic feature of an analog ROF is its programmability-how to select the rank that the filter outputs. Our filter is an analog sorting network connected to a digital rank selector. One advantage of this is that we can easily modify the circuit to simultaneously give us signals of several different ranks. Thus, even though our circuit has a moderately higher transistor count than [4], we can easily extend it to provide us a much higher level of functionality. With other contemporary ROFs, this necessitates duplication of the entire circuit.

In section 2, we describe the design of the proposed ROF. We describe the two-input sorting elements and how to construct a sorting network from these sorting elements. We then describe the digital rank selector and a simple extension to the circuit that enables us to compute the rank of a given signal. In section 3, we describe the results of some simulations of the proposed design.

## 2. DESCRIPTION OF DESIGN

In this section, the design of the proposed ROF will be described. First we describe the basic two-input sorting input. This is followed by the description of the scheme for arranging the two-input elements to make the complete network. Then we describe the digital rank selector and the scheme to extract the rank of a given signal from the input.

### 2.1 Two-input Sorting Elements

The two input sorting element uses a CMOS differential amplifier setup connected to two "voltage choosing" MOSFET configurations. Each of these basically functions as an analog multiplexer, with one MOSFET in saturation and the other in cutoff. The circuit diagram is shown in Figure 1. Similarly, nine transistors will be needed for every sorting element used. $V_1$ and $V_2$ are the two input ports and the output is tapped out of *min* and *max*.

When $V_1 > V_2$, the voltage output of the differential amplifier drops, and establishes a low impedance path between $V_2$ and *min*. Also, the MOSFET between *min* and $V_1$ is driven into cutoff, thus mutually isolating the two nodes. Simultaneously, a low impedance connection forms between $V_1$ and *max*, and a similar high impedance forms between *max* and $V_2$. When $V_1 < V_2$, the opposite process occurs, making *min* go to $V_1$ and *max* become equal to $V_2$.

### 2.2 Connecting Sorting Elements to Form Sorting Network

The two-input sorting elements may be connected in any valid sorting network configuration to produce a valid ROF. In Figure 2 we present a valid configuration for a four-input device [8]. Any other valid configuration has to use at least five comparators.

An alternative configuration is possible, where the sorting elements are arranged as a selection network. A selection network has a minimum comparator count of $O(n)$ for an $n$-input circuit, as opposed to an $O(n \log n)$ for a corresponding sorting network. Thus a selection network would have a smaller transistor count but lower programmability and other options available to circuit at run time.

### 2.3 Digital Rank Selector

We now describe the digital rank selector that enables us to program the circuit at run time. For an $n$-input ROF, the rank selector consists of $n$NMOS transistors connected to a common output terminal, as shown in Figure 3. The selection bits are connected to the gates of the MOSFETs.

A voltage of $V_{cc} = 5V$ applied to one of these bits selects the rank. The other bits are held at $V_{ee} = -5$ V.

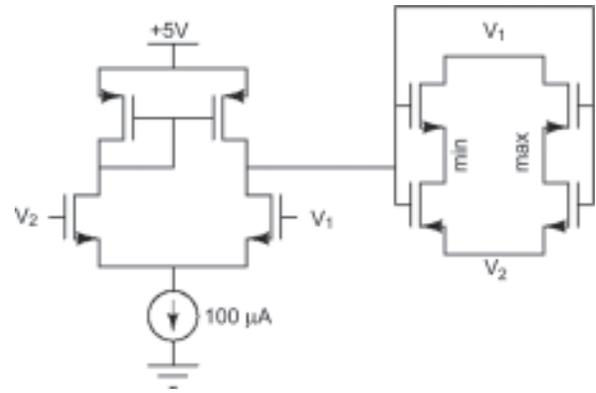

Figure 1: Two-input sorting element.

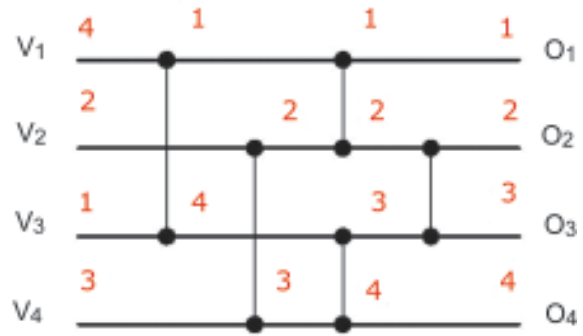

Figure 2: Two-input sorting elements forming sorting network *(circuit switching when the inputs are in the order shown in red)*.

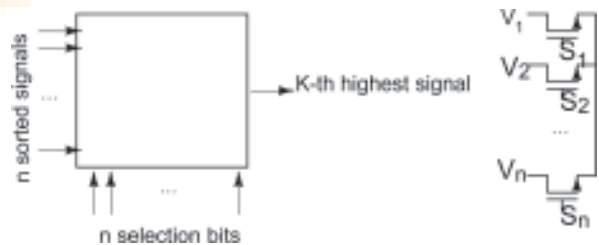

Figure 3: Selector circuit.

### 2.4 Extracting Rank of Given Signal

Observe the circuit in Figure 2, operating in reverse. The operation of the circuit elements is decided by whether the downstream signals in the original circuit need to be sent out immediately or crossed before proceeding. If we view the signals as flowing backwards through the circuit, we see that at each input, we get the rank of the corresponding signal [Figure 4].

This forms the key to the reverse operation of this circuit. The simplicity by which this extension can be achieved is due to the modularity of the design. To identify the rank of a given signal we produce another output corresponding to each input signal. For each rank, we accept a metadata signal; say one volt for the least output, 1.5 V for the second rank, etc. Adjacent to each input, we produce a one volt voltage if it is currently the least signal entering the circuit, 1.5 V if it is the signal of the second rank, etc. We have to extend the sorting element to allow for "backward" flow of this metadata. This is described in Figure 5. This is similar to information flowing upstream through the circuit -when the first input to the sorting element is the minimum of the two, these meta-signals flow straight through. When the second input to the sorting element is the minimum, the signals at the inputs "cross" each other. We do the same with the metadata - they are crossed to reflect the current state of the sorting element.

When metadata signals are also connected in the same sorting network configuration, the output at each of the input signals is representative of the rank of that signal, as shown in Figure 6.

### 2.5 Overview of Whole Design

The ROF therefore consists of a sorting network which pumps its outputs into a rank-selector. Optionally, we have metadata flowing upstream through the circuit. This is shown in Figure 7.

### 3. RESULTS

This section discusses some of the simulation results done to validate the proposed design. We report the results of simulating a four-input sorting network with LTspice/SwitcherCAD III [9]. Six sorting elements are placed in a selection sort configuration. Sinusoidal pulses from one to eight kHz and four volt p-p are given at each of the inputs. The input waveforms are shown in Figure 8a and the output as simulated by SPICE is shown in Figure 8b.

### 4. CONCLUSION

We have discussed a simple implementation of an analog ROF using CMOS voltage-mode two input sorters. These elements have been arranged as a sorting network to form the final ROF. The main advantage of our design is modular design. This resulted in a device that could be easily extended to perform other related activities such as calculating the rank of a given signal. Also, extracting signals of multiple ranks simultaneously leads to significant improvement in the transistor count compared to other designs.

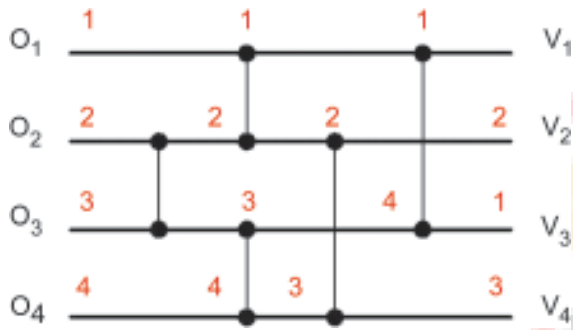

Figure 4: Mirror image of Figure 2.

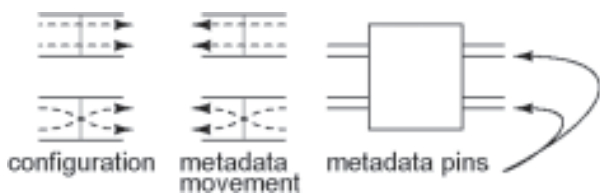

Figure 5: Upstream information flow through sorting element.

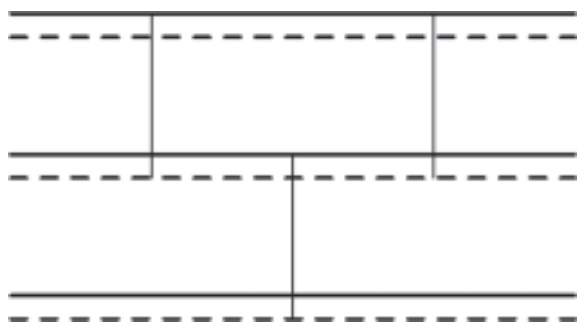

Figure 6: Determining rank of given input signal.

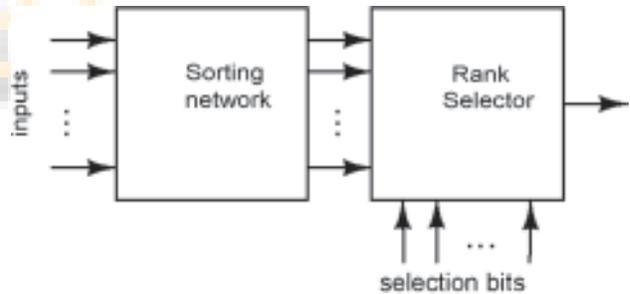

Figure 7: Block-level description of design.

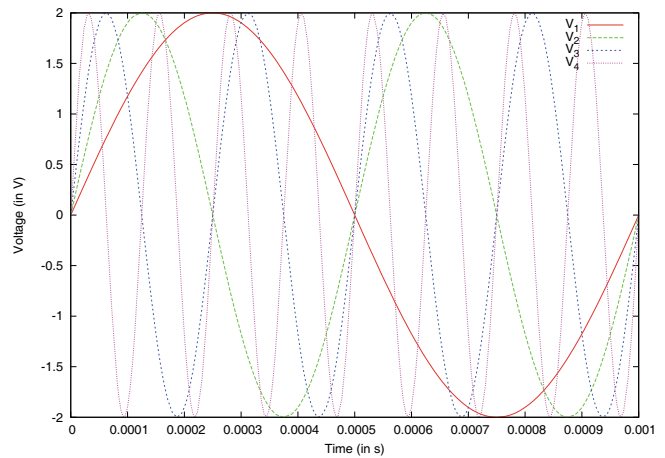

Figure 8a: Experimental results - Inputs.

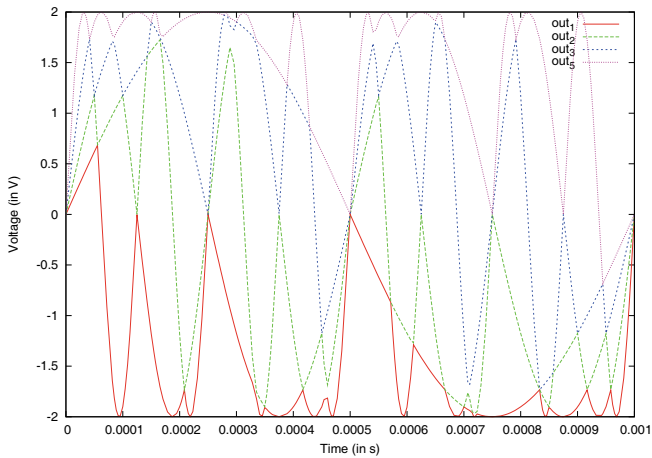

**Figure 8b:** Experimental results - Outputs.

### AUTHORS

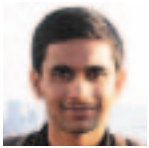

**Mukund R** is a final year undergraduate student of the department of Computer Science and Engineering at the Indian Institute of Technology, Guwahati. His research interests include theoretical computer science and programing languages. He also enjoys working on problems in the fields of electronics and mathematics.

**E-mail:** mukund@iitg.ernet.in

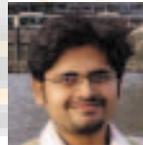

**Amit Kumar Mishra** received B.E. Degree from National Institute of Technology, Rourkela (formerly Regional Engineering College, Rourkela), in 2001; Ph.D. degree from University of Edinburgh, UK, in 2006. After completing B.E. he worked at the Defence R&D organization (DRDO) and Wipro Technologies for a year each. He has been with the ECE Department, Indian Institute of Technology Guwahati, since 2006. His research interests include analog electronics, pattern recognition, radar signal processing, VLSI DSP and ultrawideband (UWB) systems. He is a recipient of IETE (IRSI) Young Scientist award for the year 2008.

**E-mail:** akmishra@ieee.org